\documentclass[lineno]{jfm}

\usepackage{graphicx}
\usepackage{newtxtext}
\usepackage{newtxmath}
\usepackage{natbib}
\usepackage[linkcolor=black]{hyperref}
\hypersetup{colorlinks = true, urlcolor = blue, citecolor = black}
\newcommand{\RomanNumeralCaps}[1]
\linenumbers

\usepackage{soul}
\usepackage{mathtools}
\usepackage{subfig} 

\newcommand{\PF}{\mathbb{P}(\mathbf{F})}
\newcommand{\FCag}{\mathbf{F}_\textrm{\scriptsize{C}}^\textrm{\scriptsize{ag}}}
\newcommand{\forcevec}[1]{\mathbf{F}_\textrm{\scriptsize{#1}}}

\newcommand{\curled}[1]{(C_\textrm{\scriptsize{#1}})_l}
\newcommand{\proj}[1]{(P_\textrm{\scriptsize{#1}})_l}
\newcommand{\forceCag}{(F_\textrm{\scriptsize{C}}^\textrm{\scriptsize{ag}})_l}
\newcommand{\rmd}{{\rm d}}
\let\citeA\cite
\title{Solenoidal force balances in numerical dynamos}

\author{Robert J.~Teed\aff{1} and Emmanuel
  Dormy\aff{2}\corresp{\email{Robert.Teed@glasgow.ac.uk, dormy@dma.ens.fr}}}

\affiliation{
\aff{1}{School of Mathematics and Statistics, University of Glasgow, University Place, Glasgow G12 8SQ, UK}
\aff{2}{Département de Mathématiques et Applications, UMR-8553, École Normale Supérieure, CNRS, PSL
University, 75005 Paris, France}}

\begin{document}
\maketitle

\begin{abstract} %Must be single paragraph and <250 words. Currently ~200 words
 Numerical simulations of the geodynamo (and other planetary dynamos) have
 made significant progress in recent years. As computing power has
 advanced, some new models claim to be ever more appropriate
 for understanding Earth's core dynamics. One measure of the success of
 such models is the ability to replicate the expected balance between
 forces operating within the core; 
 %rotational (Coriolis) and magnetic (Lorentz)
 Coriolis and Lorentz forces are predicted to be most important. The picture is
 complicated for an incompressible flow by the existence of the pressure gradient force  which renders the gradient parts of all other forces dynamically unimportant.
 %which is not dynamically important but balances gradient parts of all other forces.
 This can confuse the situation, especially when the scale dependence of forces are considered.  
 In this work we investigate force balances through the alternative
 approach of eliminating gradient parts of each force to form `solenoidal force balances'.
 We perform a lengthscale dependent analysis for several spherical simulations
 %in different regimes 
 and find that removal of gradient parts 
 %through curled or projected forces
 offers an alternative
 %different and clearer 
 picture of the force balance compared to looking at traditional forces alone.
 Solenoidal force balances provide some agreement with the results of previous studies 
 %at large scales
 but also significant differences. They offer a cleaner overall picture of the dynamics and introduce differences at smaller scales.
 This has implications for geodynamo models purporting to have reached Earth-like regimes: in order to achieve a meaningful comparison of forces, only the solenoidal part of forces should be considered.
\end{abstract}


\section{Introduction}
\label{sec:intro}

The geomagnetic field is widely believed to be generated and maintained by dynamo action via convective fluid motion within Earth's iron-rich outer core. Numerical simulations of the geodynamo (and other planetary dynamos) are a vital component for enhancing our understanding of core dynamics and the dynamo process. %Such simulations have made great advances in recent years as computing power has improved. 
%There now exist plenty of published studies drawing in vast numbers of simulations that are performed across a range of input parameters \citep{christensen1999} {\bf (**MORE CITATIONS**)}. 
As computing resources improve, these studies have typically responded by moving input parameters of simulations closer to those of Earth's core.
%(e.g.~by lowering viscosity as much as numerically possible). 
This has led to claims by authors that each new generation of simulations are more appropriate models for the geodynamo \citep{yadav2016,aubert2017,schaeffer2017,gillet2021}. Despite this, simulations still remain far from the geophysically accurate parameter regime; the viscosity, in particular, remains hugely enhanced in all models. For this reason many studies attempt to draw conclusions for the geodynamo by looking for trends in the numerical results, 
making the (implicit) assumption that the various dynamo state(s) produced in simulations are appropriate for the geodynamo. 

In this context it is therefore vital that scaling laws and comparisons with observations are made using simulations relevant to Earth's core dynamics and the geomagnetic field. 
A possible measure is to consider the strength and balance of forces in simulations and compare these with the expected balances in the core itself.
Within simulations, the balance of forces is dependent on spatial scale (and location) and varies with the regime of the saturated state. The leading order force balance 
observed in numerical models
is usually cited to be `quasi-geostrophic' (QG); namely balance between the Coriolis force and the pressure gradient. A secondary balance then exists between Lorentz (`magnetic', M), buoyancy (`Archimedean', A), and Coriolis (C) - the so-called `MAC-balance' - as expected in the geodynamo, with remaining forces (inertia and viscous) playing weaker roles 
\citep{schwaiger2019,schwaiger2021}.
However, the gradient parts of all forces play no role in the dynamics of the incompressible (Boussinesq) flow. 
One approach motivated by a desire to analyse the Coriolis force as a contributor in the MAC-balance instead of the QG-balance, is to subtract the pressure gradient from the Coriolis force to form a so-called `ageostrophic Coriolis force'.
However, this quantity (along with all other forces) retains a gradient part, which could be an unsatisfactory side-effect of forming such a simple construction.
An alternative approach, which we advocate in this work, is to 
focus on the dynamically relevant effects and
form a `{\em solenoidal} balance of forces'. 
This is achieved by eliminating all gradient parts of the force balance; we discuss two methods for this in Section \ref{sec:forcedecomp}. 
The main aims of this paper are two-fold: 1) to introduce approaches for considering the solenoidal balance of forces in spherical dynamo simulations; 2) to examine the solenoidal force balance on a few selected cases.

\section{Mathematical formulation}

\subsection{Mathematical and numerical setup}

We consider a spherical shell, in spherical coordinates: $(r,\theta,\phi)$, filled with a Boussinesq, electrically conducting fluid of constant density, $\rho$, and located between $r=r_\textrm{i}$ and $r=r_\textrm{o}>r_\textrm{i}$. The shell rotates with rate $\mathbf{\Omega}=\Omega \,\mathbf{\hat{z}}$ where $z$ is the vertical coordinate. Gravity acts radially inwards such that $\mathbf{g}=-g\mathbf{r}$. The intrinsic diffusivity parameters (the kinematic viscosity, $\nu$, thermal diffusivity, $\kappa$, and magnetic diffusivity, $\eta$) of the fluid are assumed to be constant.
At both boundaries,
%($r=r_\textrm{i}$ and $r=r_\textrm{o}$),
we use impenetrable, rigid, electrically insulating, isothermal conditions with 
a temperature difference of $\Delta T$ maintained between the boundaries, 
allowing differential heating to drive convection.
The coupled set of partial differential equations governing the evolution of velocity, $\mathbf{u}$, pressure, $p$, temperature, $T$, and magnetic induction, $\mathbf{B}$, are
\begin{subequations}
\label{eq:gov}
\begin{gather}
 E_\eta\left(\frac{\partial\mathbf{u}}{\partial t} + \mathbf{u}\cdot\nabla\mathbf{u}\right) = -\nabla p - 2\, \mathbf{\hat{z}}\times\mathbf{u} + (\nabla\times\mathbf{B})\times\mathbf{B} + \widetilde{Ra}\, T\, \mathbf{r} + E\, \nabla^2\mathbf{u}\,, \label{eq:NSeq} \tag{\theequation $a$}\\
 \frac{\partial T}{\partial t} + \mathbf{u}\cdot\nabla T = q\, \nabla^2T\,, \qquad \qquad
 \frac{\partial\mathbf{B}}{\partial t} - \nabla\times(\mathbf{u}\times\mathbf{B}) = \nabla^2\mathbf{B}, \label{eq:indeq}\tag{\theequation $b,c$} \\
 \nabla\cdot\mathbf{u} = 0\,, \qquad
 \nabla\cdot\mathbf{B} = 0\,,\tag{\theequation $d,e$}
\end{gather}
\end{subequations}
where we have nondimensionalised using lengthscale, $d=r_\textrm{o}-r_\textrm{i}$, timescale, $d^2/\eta$, temperature scale, $\Delta T$, and magnetic scale $\sqrt{\rho\mu_0\eta\Omega}$.
The nondimensional parameters appearing in our set of equations are the magnetic Ekman number, $E_\eta$, the modified Rayleigh number, $\widetilde{Ra}$, Ekman number, $E$, Roberts number, $q$,  defined as
\begin{subequations}
\begin{equation}
 E_\eta=\frac{\eta}{\Omega d^2}\, ,
 \qquad \widetilde{Ra}=\frac{\alpha g\Delta T d}{\Omega\eta}\, , 
 \qquad E=\frac{\nu}{\Omega d^2}\, , 
 \qquad q=\frac{\kappa}{\eta}\, .
 \tag{\theequation $a$--$d$}
\end{equation}
\end{subequations}
The Rayleigh number we use is a modified version of the classical Rayleigh number, $Ra$,
where $\widetilde{Ra}=Ra\,E\,q$. The aspect ratio of the shell, $r_\textrm{i}/r_\textrm{o}$, is set to the Earth-like value of 0.35. 
Alternative input parameters (sometimes used in dynamo simulations) are the 
Prandtl number, $Pr\equiv E/qE_\eta =\nu/\kappa$ and the
magnetic Prandtl number, $Pm\equiv E/E_\eta=\nu/\eta \, .$

We numerically solve the governing equations, (\ref{eq:gov}$a$--$e$), using the Leeds spherical dynamo code \citep{willis2007}. 
The output parameters we report for each dynamo simulation are the magnetic Reynolds number, $Rm$, the Elsasser number, $\Lambda$, the modified Elsasser number \citep{dormy2016}, $\Lambda'$, 
which are defined by
\begin{subequations}
\begin{equation}
 Rm=\frac{Ud}{\eta}\, , \qquad \Lambda=\frac{B^2}{\rho\mu_0\eta\Omega}\, , \qquad \Lambda'=\Lambda\frac{d}{Rm\ell_B}=\frac{B^2}{\rho\mu_0\Omega U\ell_B}\, ,
 %\quad f_\textrm{dip}=, \quad f_\nu=,
 \tag{\theequation $a,b,c$}
\end{equation}
\end{subequations}
where $U$ and $B$ are (dimensional) rms values of the velocity and magnetic field, respectively, and $\ell_B^2=\int_V{\mathbf{B}^2 \,\rmd V}/\int_V{(\nabla\times\mathbf{B})^2 \, \rmd V}$ is a measure of the typical magnetic dissipation lengthscale. 
We also report the dipolarity, $f_\textrm{dip}$ \citep{teed2014}, and the ratio of the energy being dissipated by viscous forces to the total energy dissipation (viscous and ohmic), $f_\nu$ \citep{dormy2018}.
All quantities are averaged over space and time.

\subsection{Forces and solenoidal forces}
\label{sec:forcedecomp}

Forces in our model can be identified from (\ref{eq:NSeq}). They are the inertial, $\forcevec{I}$, pressure gradient, $\forcevec{P}$, Coriolis, $\forcevec{C}$, Lorentz, $\forcevec{L}$, Archimedean, $\forcevec{A}$, and viscous, $\forcevec{V}$, forces defined by
\begin{subequations}
\label{eq:forces}
\begin{equation}
 \forcevec{I}=E_\eta\left(\frac{\partial\mathbf{u}}{\partial t} + \mathbf{u}\cdot\nabla\mathbf{u}\right)\,, \quad\quad\ 
 \forcevec{P}=-\nabla p\,, \qquad\quad
 \forcevec{C}=-2\,\mathbf{\hat{z}}\times\mathbf{u}\,, \tag{\theequation $a,b,c$}
 \end{equation}
 \begin{equation}
 \forcevec{L}=(\nabla\times\mathbf{B})\times\mathbf{B}\,, \qquad\qquad\ 
 \forcevec{A}=\widetilde{Ra}\,T\,\mathbf{r}\,, \qquad
 \forcevec{V}=E\,\nabla^2\mathbf{u}\,.\ \tag{\theequation $d,e,f$}
\end{equation}
\end{subequations}
Several recent studies \citep{aubert2017,schwaiger2019,schwaiger2021} also analyse an `ageostrophic Coriolis force' constructed by subtracting the pressure gradient from the full Coriolis force: $\FCag=\forcevec{C}-\forcevec{P}$. We will make use of shorthand notation on variables and in text: I (inertia), P (pressure), C (Coriolis), L or M (Lorentz or magnetic), A (Archimedean), and V (viscous).
A sub- or super-script `ag' refers to the ageostrophic Coriolis force.

The balances commonly spoken of in reference to Earth's core dynamics are the QG 
and MAC balances. A purely geostrophic balance leads to the Taylor-Proudman constraint, demanding $z$-independent flows. 
However, the constraint itself arises from the {\em curl} of the balance
which has eliminated gradient parts of forces. 
The constraint is thus independent of the pressure gradient and, indeed, any gradient part, which has led to attention being given to the `ageostrophic Coriolis force' defined above.
However, flows are not perfectly geostrophic and,
in fact, {\em all} forces 
listed in (\ref{eq:forces}) may have gradient parts which, combined, drive a pressure gradient.
This combined gradient force is, in an incompressible flow, immediately balanced by pressure to ensure the incompressible constraint.
As such, $\FCag$ also retains a gradient part but
none of these gradient parts are relevant to the dynamics.
A safer 
and cleaner mathematical
approach
is to directly eliminate the gradient part of the Coriolis term and to split the contribution to pressure driven by each force separately. This allows access to the remnant parts (and their balance), which can be thought of as versions of each force with the gradient part removed; we refer to these as `solenoidal forces'.

\cite{soderlund2015,yadav2016,AurnouKing2017}
used volume integrated forces to analyse the competition of Coriolis and Lorentz forces.  \cite{aubert2017} expanded on this approach by considering scale dependence. A systematic study using this approach was performed by \cite{schwaiger2019}. They later focussed on extraction of potentially relevant lengthscales from force balances \citep{schwaiger2021}.
In this work, 
we go further by
considering two methods of eradicating gradient parts of the forces to form solenoidal forces: forming the curl of each force; and projecting each force onto its solenodial (i.e.~non-gradient) part.

\subsubsection{Decomposition of forces}
\label{sec:forces}

Using a known approach \citep{aubert2017,schwaiger2019}, each force defined in (\ref{eq:forces}) is written as a combination of scalar potentials ($\mathcal{R}$, $\mathcal{S}$, and $\mathcal{T}$), which, in turn, are expanded in spherical harmonics of degree $l$ and order $m$. (Partial) integration of the force vector over the volume (see the Supplementary Material) allows us to define
\begin{equation}
F_l^2={\cal F}(\mathcal{R},\mathcal{S},\mathcal{T})\equiv 2\sum_{\mathclap{m=0}}^l\hspace{-1mm}{\vphantom{\sum}}' \int_{r_\textrm{i}+b}^{r_\textrm{o}-b} \left[|\mathcal{R}_l^m|^2 + l(l+1)\left(|\mathcal{S}_l^m|^2 + |\mathcal{T}_l^m|^2\right)\right] \, r^2 \, \rmd r\, ,
\end{equation}
which gives the power spectra of an individual {\it force} as a function of $l$. Here, the prime denotes a halving of the $m=0$ term and $b\sim O(E^{1/2})$ represents the boundary layer thickness.

%%%%%%%%%%%
\subsubsection{Decomposition of curl of forces}
A straightforward way to form a solenoidal representation of forces is to consider the curl of each force. This approach was followed by \cite{dormy2016,schaeffer2017}. We curl each force in (\ref{eq:forces}) 
and then write them as a combination of scalar potentials ($\hat{\mathcal{R}}$, $\hat{\mathcal{S}}$, and $\hat{\mathcal{T}}$). In an identical process to section \ref{sec:forces} (see the Supplementary Material), this allows us to define 
$C_l^2={\cal F}(\hat{\mathcal{R}},\hat{\mathcal{S}},\hat{\mathcal{T}})$
which gives the power spectra of an individual {\it curl of a force} as a function of $l$.

Such spectra have a natural tendency to peak at smaller scales and thus need sufficient resolution \citep[see][]{hughes2019}.
It is easy to compensate spectra for this increase at smaller scale. Since the double curl of a solenoidal force relates to its Laplacian, we chose to compensate with $\hat{C}_l=(l(l+1))^{-1/2}C_l$ (though very similar plots would have been obtained though a compensation via $l^{-1}$). It should be stressed that the compensated spectra, of course, present exactly the same crossings between curves as the uncompensated spectra.

The curl of forces offers a unique representation of the solenoidal part of forces. As we shall see below, this uniqueness does not apply to projected forces. In particular, these compensated spectra should not be confused with an `uncurl' which would have to involve the radial lengthscale as well as, for a solenoidal uncurl, the gradient of an arbitrary harmonic potential.

\subsubsection{Formation and decomposition of projected forces}

An alternative, more formal, procedure for forming solenoidal forces 
is to calculate the projection of forces onto their solenoidal part. 
This projection stems from the Helmholtz-Hodge decomposition, namely the fact that a smooth vector field 
in a bounded domain can be uniquely decomposed into a pure gradient field and a divergence-free vector parallel to the boundary {(at the boundary)}.
This results in the so-called Leray projector; for a given force, $\mathbf{F}$, if we write $\mathbf{F}=\nabla\times\mathbf{A}+\nabla\varphi$ for some potentials $\mathbf{A}$ and $\varphi$, then $\nabla\cdot\mathbf{F}=\nabla^2\varphi$. Given boundary conditions, this can be solved for $\varphi$, whence the projected force is defined as 
$\PF=\mathbf{F}-\nabla\varphi=
\mathbf{F}-\nabla\left((\nabla^2)^{-1}\nabla\cdot\mathbf{F}\right)
$.

Such an approach is often used when computing incompressible flows and is then referred to as the Temam-Chorin algorithm.
An impermeable boundary imposes $\mathbf{u}\cdot \mathbf{n}=0$ on the boundary of the domain and  the flow (or the sum of all forces) can thus be uniquely projected. In other words, the potential $\varphi$ is uniquely determined with $\partial_n\varphi=\mathbf{F}\cdot \mathbf{n}$ at the boundary. When considering individual forces, the fact that one wants to reconstruct a divergence-free vector field, which has no physical reason to be parallel to the boundary raises the issue of non-uniqueness. Indeed the potential $\varphi$ is then determined up to an arbitrary harmonic potential set by the boundary conditions imposed on the projected field (as would be the case for an `uncurl', see above). 
Here we imposed that each force is projected onto the space of divergence-free vector functions having vanishing normal
component along the boundary. This corresponds to the Leray projector; it is a sensible choice, but not imposed by physics. 
A similar projection was considered by \citeA{hughes2019} in Cartesian geometry simulations though without specifying the harmonic field.

As before, $\PF$ is then written as a combination of scalar potentials ($\tilde{\mathcal{R}}$, $\tilde{\mathcal{S}}$, and $\tilde{\mathcal{T}}$), which, following the same method of section \ref{sec:forces}, allows us to define 
$P_l^2={\cal F}(\tilde{\mathcal{R}},\tilde{\mathcal{S}},\tilde{\mathcal{T}})$
 giving the power spectra of an individual {\it force projection} as a function of $l$.

\section{Results}
We illustrate here the effect of considering the solenoidal force balance on a few simple cases which 
do not claim to be state of the art, but which highlight the effects of dropping the gradient part of all forces. We make use of regime diagrams \citep{christensen1999,dormy2016,moffatt2019}.
Tab.~\ref{tab:sims} lists input and output parameters of dynamo simulations included in this study; here the criticality of the Rayleigh number is compared to the onset of (non-magnetic) convection \citep{dormy2004}.

\begin{table}
    \centering
    \begin{tabular}{c|ccccc|ccccccc}
    Run & $E$ & $E_\eta$ & $q$ & $Pm$ & $\frac{\widetilde{Ra}}{\widetilde{Ra}_c}$ & $Rm$ & $Ro_\ell$ &  $\Lambda$ & $\ell_B$ & $\Lambda'$ & $f_\textrm{dip}$ & $f_\nu$ \\
         %Run & $E$ & $Ra/Ra_c$ & $Pm$ & $Rm$ & $\Lambda$ & $f_{dip}$ & $E_K$ & $E_M$  \\
         \hline
%         HD & $10^{-4}$ & - &	- &	- &	10 & 163.204 &	0.0181 & 0	& - &	0	& 0 &	1 \\
         A & $10^{-4}$ & 8.3$\times10^{-6}$ & 12 & 12 &	2.05 & 206.537 &  0.00417 &  0.237 & 0.101 & 0.0113 & 0.938 & 0.974 \\
        % \WDplus & $10^{-4}$ & $10^{-4}$ & 1 & 1 & 10 & 79.165 & 0.0388 &  2.458 & 0.0834 & 0.372 & 0.954 & 0.461 \\
         B & $10^{-4}$ & 8.3$\times10^{-6}$ & 12 & 12 & 10 & 863.002  &	0.0313 &  81.772 & 0.0347 & 2.731 & 0.831 & 0.422 \\
         C & $10^{-4}$ & $10^{-4}$ & 1 & 1 & 30 & 293.419 & 0.152 & 0.0553 & 0.823 &  0.0892 & 0.317 & 0.871\\
         D & $10^{-5}$ & $10^{-5}$ & 1 & 1 & 10 & 149.624 & 0.0127 & 2.116 &  0.0513 & 0.276  & 0.908 & 0.288 \\
    \end{tabular}
    \caption{Input and output parameters of dynamo simulations performed for this study.
    }
    \label{tab:sims}
\end{table}

\subsection{Hydrodynamic solution}

\begin{figure}
\centerline{
(a)\hskip -5mm{\label{fig:HD_FlnoBL}\includegraphics[width=0.33\linewidth]{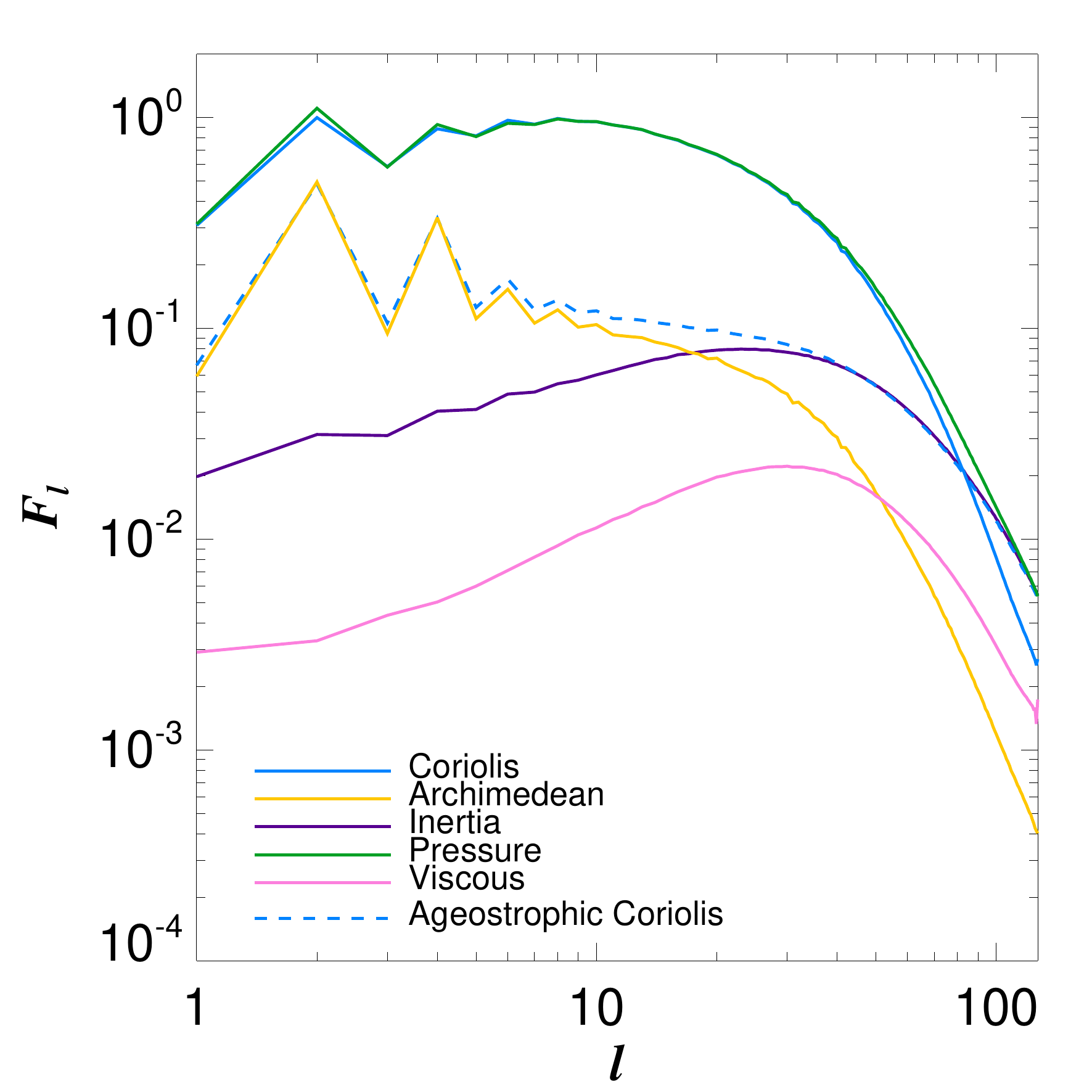}}
\hspace{1mm}
(b)\hskip -5mm{\label{fig:HD_ClnoBL}\includegraphics[width=0.33\linewidth]{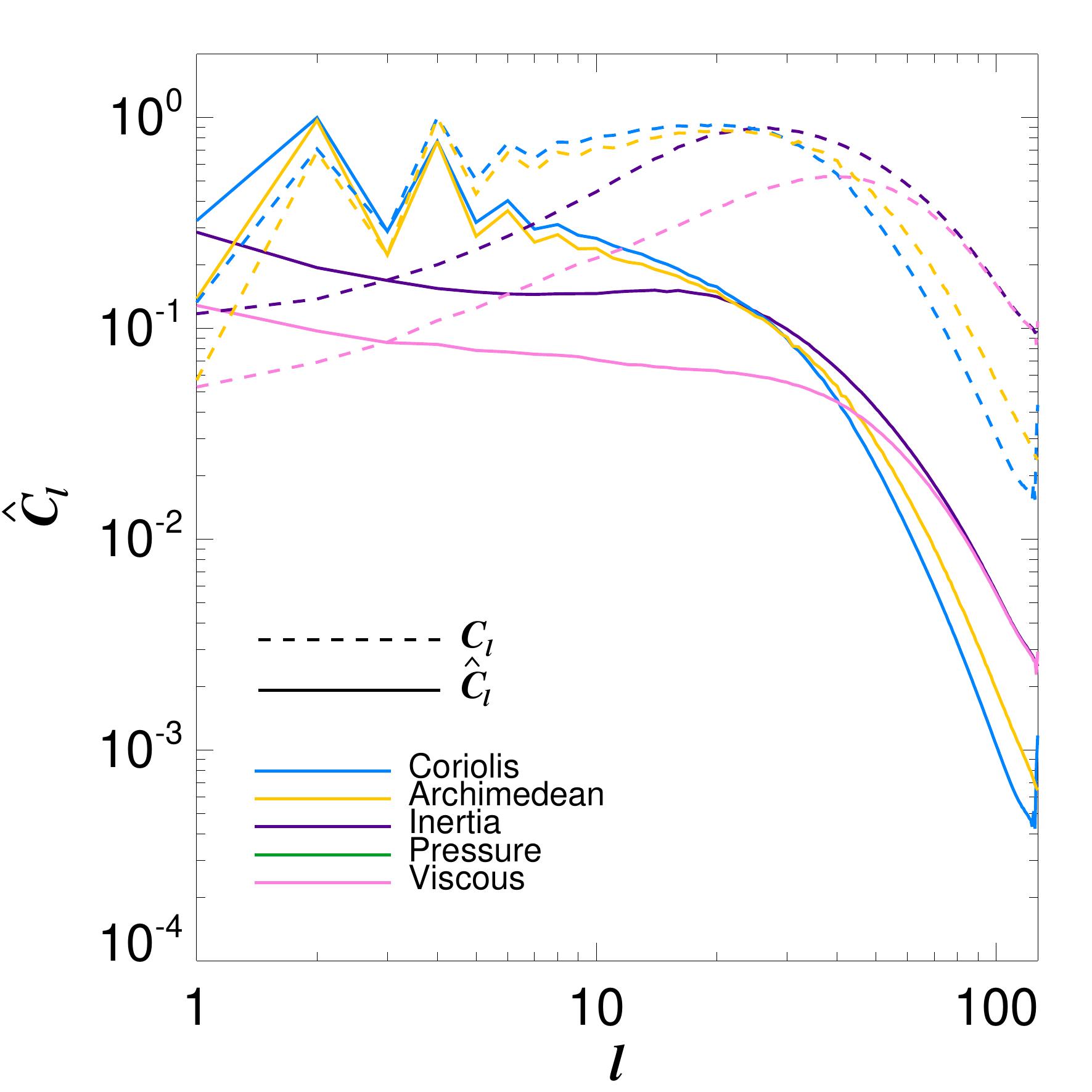}}
\hspace{1mm}
(c)\hskip -5mm{\label{fig:HD_PlnoBL}\includegraphics[width=0.33\linewidth]{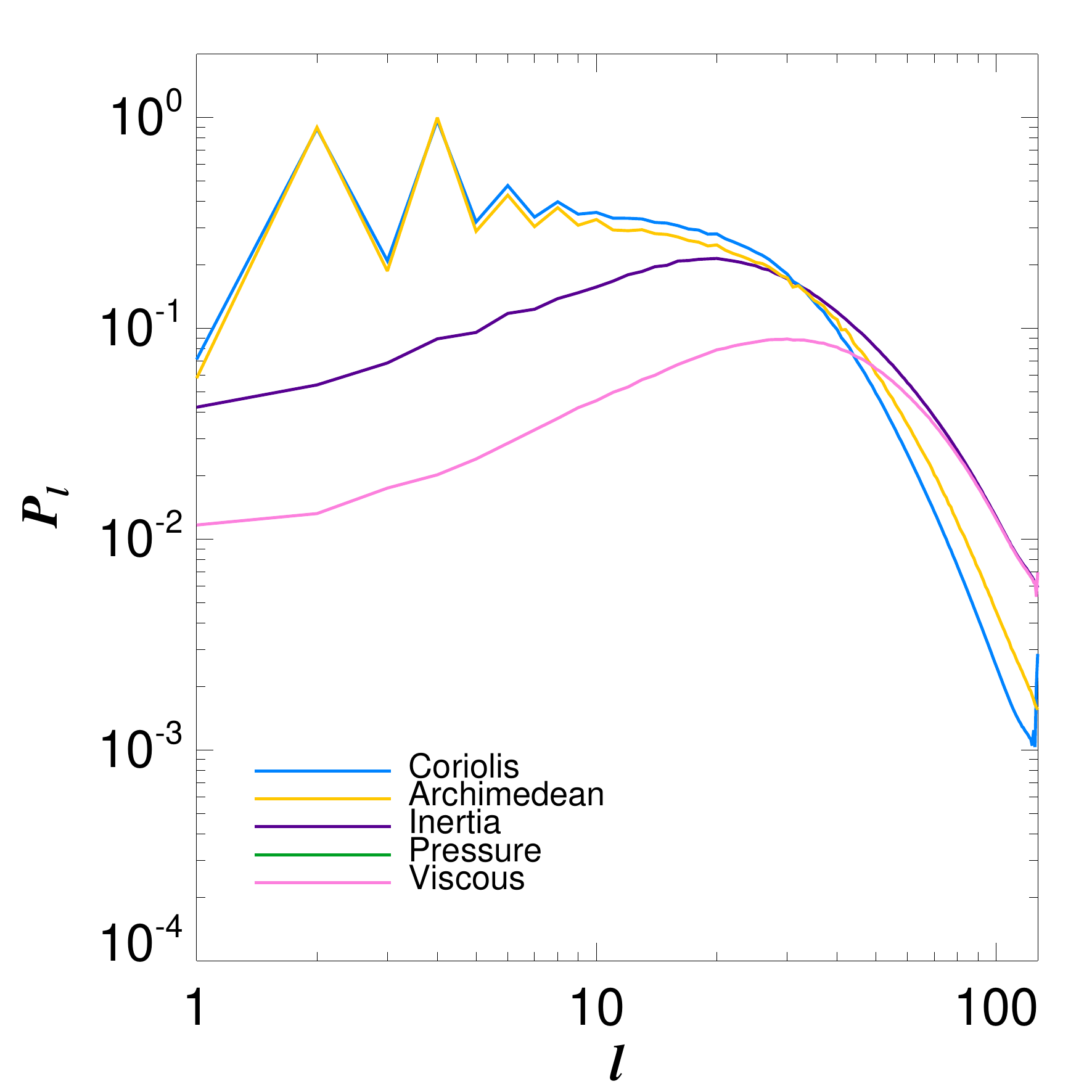}}}
\caption{Comparison of representation of forces for the HD run ($E=10^{-4}$, $Ra=10Ra_c$). (a) forces, (b) curl of forces (compensated and uncompensated) and (c) projected force. All quantities have been averaged in time and space, yet boundary layers have been removed.
}
\label{fig:HD10}
\end{figure}

We start by considering a purely hydrodynamic (HD) simulation with no magnetic field.
In that case the viscous timescale $d^2/\nu$ is used to non-dimensionalise the system (as opposed to $d^2/\eta$ elsewhere). The relevant parameters are $E=10^{-4}$,
$Pr=\nu/\kappa=1$, and $Ra/Ra_c= 10$.

\begin{figure}
\centerline{
{\label{fig:HD_mer_ur}\includegraphics[width=0.18\linewidth]{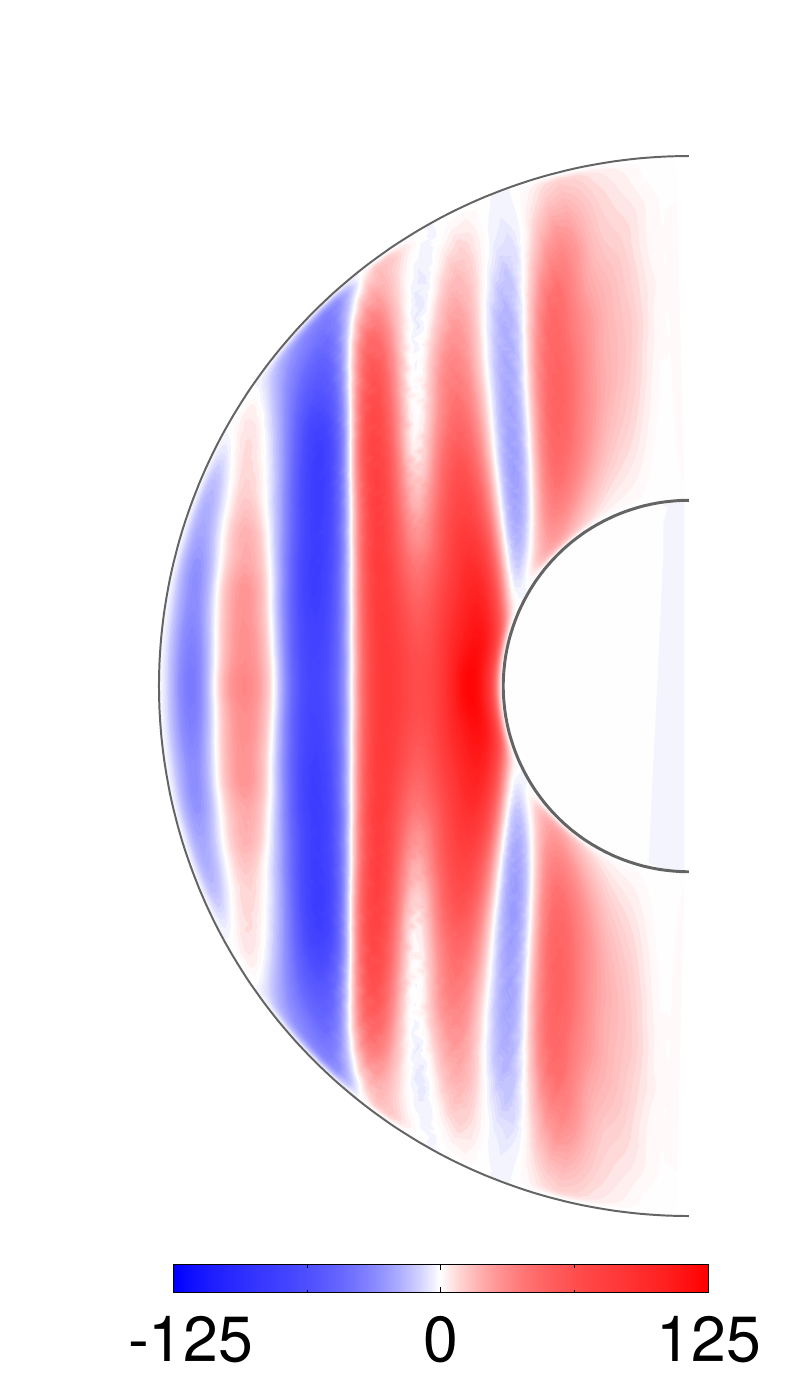}}
\hspace{8mm}
{\label{fig:WD_mer_ur}\includegraphics[width=0.18\linewidth]{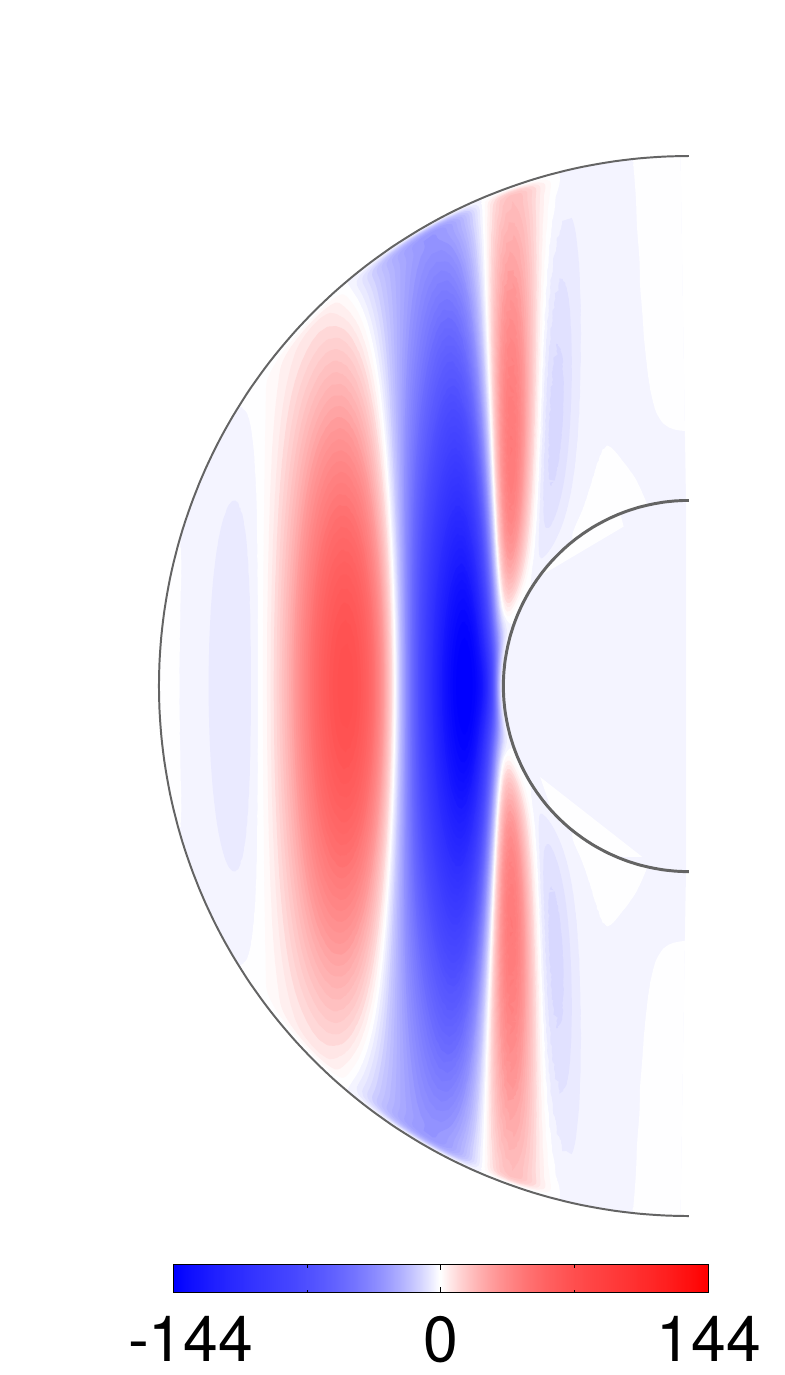}}
\hspace{8mm}
{\label{fig:SD_mer_ur}\includegraphics[width=0.18\linewidth]{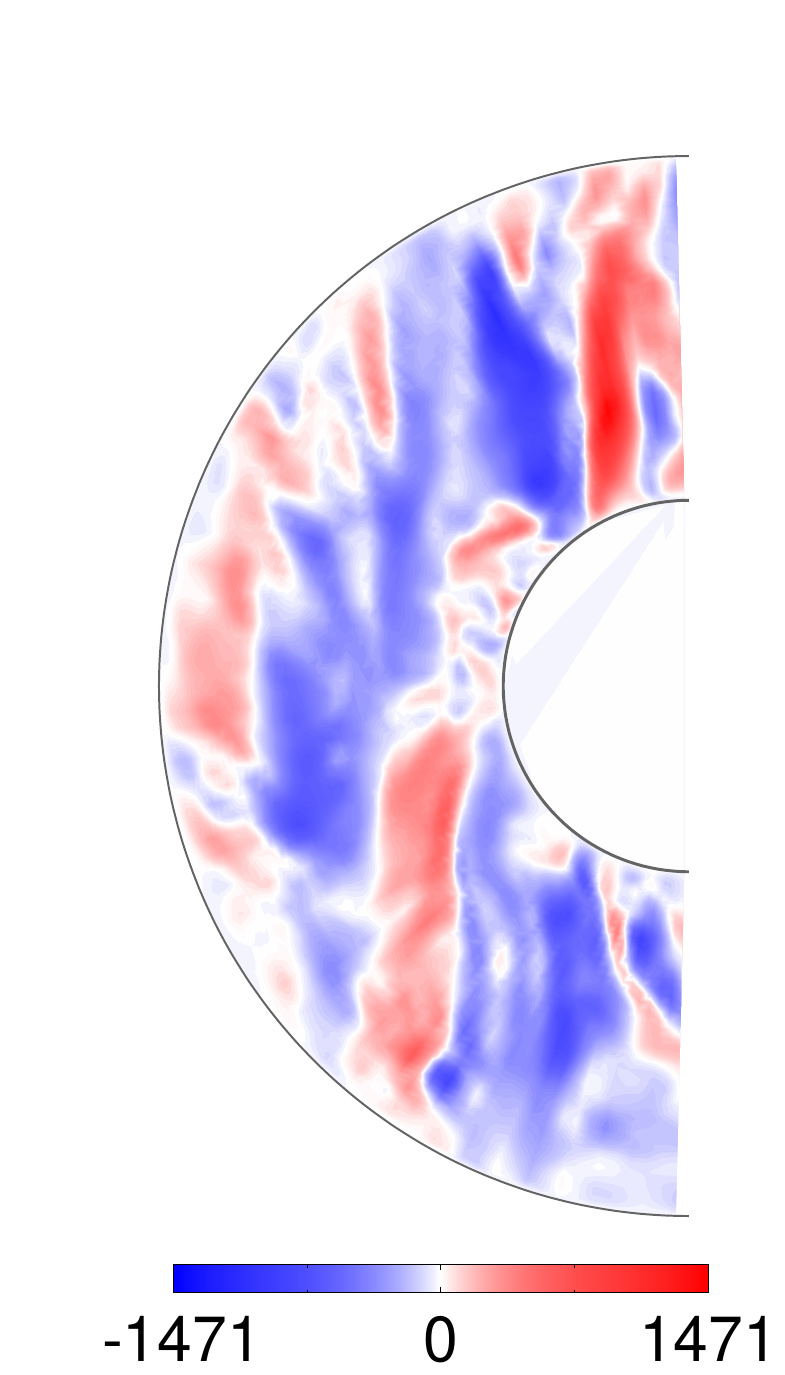}}}
\centerline{\hskip 1.0cm (a) HD \hskip 2.2cm (b) Case A \hskip 1.9cm (c)
 Case B}
\caption{Meridional sections of $u_r$ in different regimes.}
\label{fig:mersecs10}
\end{figure}

In Fig.~\ref{fig:HD10}a,b,c we present spectra plots for $F_l$, $\hat{C}_l$, and $P_l$ respectively for the HD run.
The plots of $F_l$ show a dominant geostrophic balance across most scales; 
we refer to balances involving the pressure gradient as zeroth order balances. A first order AC\textsuperscript{ag} balance exists at large scales while remaining forces are sub-dominant. At small scales the geostrophic balance is broken by inertia, which enters the zeroth order balance, replacing the Coriolis force. The viscous force appears negligible at all lengthscales.
In Fig.~\ref{fig:mersecs10}a, we plot a meridional section of $u_r$, which confirms the near-geostrophy; flow structures are predominantly independent of rotation axis.
In solenoidal force plots (Figs.~\ref{fig:HD10}b,c) the necessary absence of the pressure gradient (and hence, geostrophic balance) better reveals the balance controlling the dynamics. A first order AC balance is seen at large lengthscales, equivalent to that of AC\textsuperscript{ag} in $F_l$. Strikingly, however, the viscous force shows increased significance at smaller lengthscales effecting a transition in first order balance from AC to VI for $l\gtrsim 40$.
This important feature is lost in the plots of $F_l$ because of the dominance of the zeroth order balance combined with the assumption that only the Coriolis force balances the pressure gradient.

Fig.~\ref{fig:HD10}b also shows a comparison between uncompensated curled forces, $C_l$, in dashed lines, and their compensated counterpart $\hat{C}_l$ in solid lines. As expected, $\hat{C}_l$ partly compensates for the power introduced at smaller scales by the extra spatial gradient. In all remaining plots we use $\hat{C}_l$ for better comparison with $F_l$ and $P_l$.

It is interesting to compare the ageostrophic Coriolis force with $\curled{C}$ and $\proj{C}$.
The deviation of $\forceCag$ in Fig.~\ref{fig:HD10}a from its solenoidal counterparts in Fig.~\ref{fig:HD10}b,c demonstrates the contribution of the gradient parts of remaining forces.
The effect is most pronounced at smaller scales where 
the geostrophic balance is broken by the inertial force.

Fig.~\ref{fig:HD10} neatly shows the misinterpretation that is possible 
by
retaining gradient parts of forces.
Solenoidal forces provide a more 
precise
version of the balance by removing these complications.

\subsection{Dynamo solutions}

The presence of the Lorentz force in dynamo simulations adds to the available force balances.
It has long been proposed that `weak field' (with $\forcevec{L}\ll\forcevec{C}$) and `strong field' (with $\forcevec{L}\sim\forcevec{C}$) regimes are distinct dipolar dynamo solutions on separate regime branches \citep[see][for the history of these ideas]{moffatt2019}.
\cite{schwaiger2019} introduced the term `QG-Hybrid' for regimes close to the onset of dynamo action where the Lorentz force is weak.
A change in regime can be identified by a sharp change in the behaviour of the solution for a small change in input parameter, typically $Ra$. For example, this could be whether or not the Lorentz force enters a leading order balance at a range of lengthscales.
Previous studies \citep{christensen1999,dormy2016} guide us on suitable input parameters for obtaining various
regimes.
%To observe a direct transition between WD and SD, a small value of $E_\eta$ is required \citep{dormy2016}. 
At $E_\eta=8.3\times10^{-6}$ and $E=10^{-4}$ (i.e. $Pm=12$), we observe a discontinuous jump between branches as $\widetilde{Ra}$ is varied (we have not yet studied bistability).
For fixed values of $E_\eta$ and $E$, this allows us to consider: a regime with weak forcing that is a `weak field' 
\citep{dormy2016}
or `QG-Hybrid' \citep{schwaiger2019} solution (case A); a regime with stronger forcing where the Lorentz force is important (case B). Case B is likely more typical of the solutions commonly presented in previous literature.
Although less relevant for the geodynamo, 
for completeness, we also discuss a fluctuating multipolar 
%(FM) 
solution (case C).

\begin{figure}
\centerline{
(a)\hskip -5mm{\label{fig:Pm12_2.05Rac_FlnoBL}\includegraphics[width=0.33\linewidth]{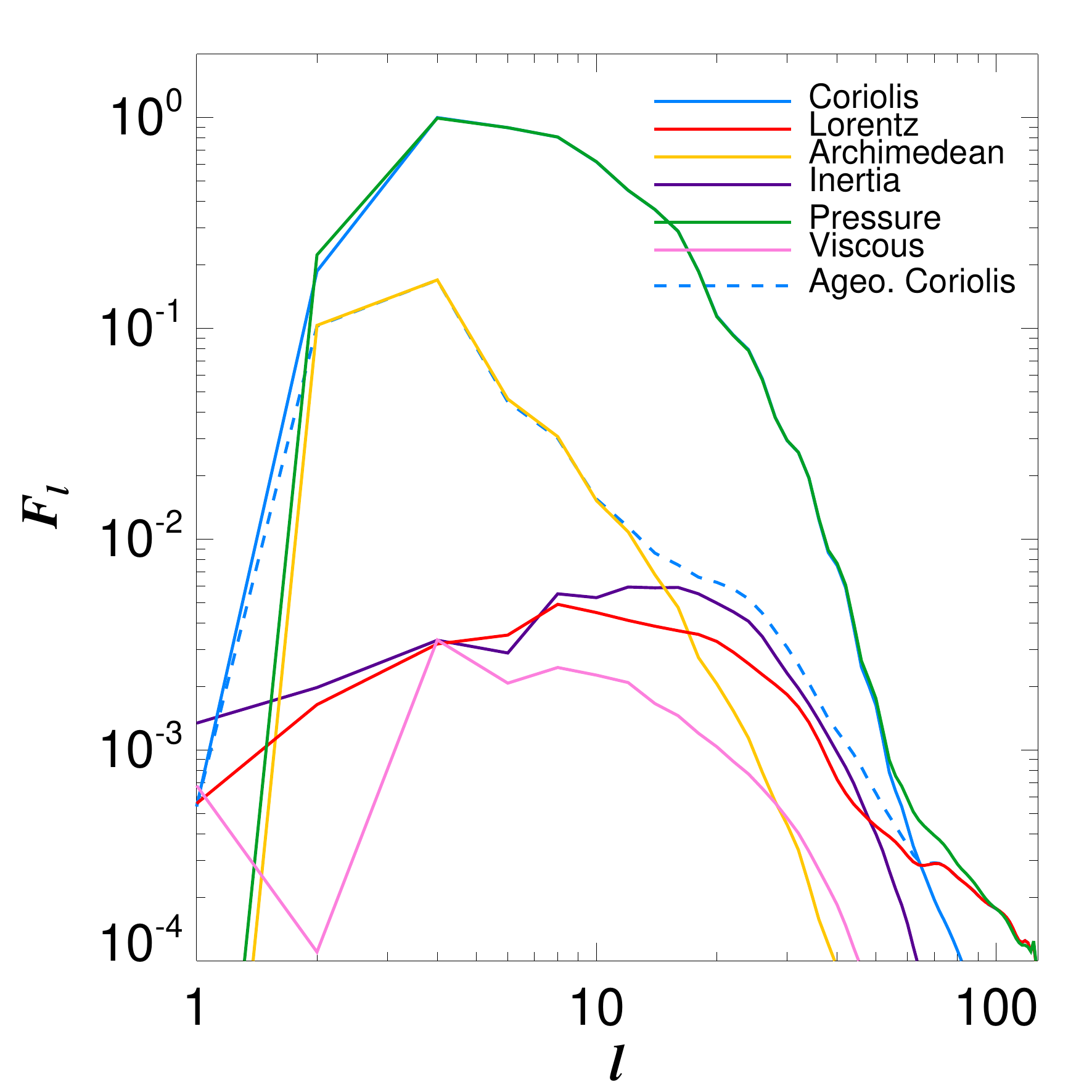}}
\hspace{1mm}
(b)\hskip -5mm{\label{fig:Pm12_2.05Rac_ClnoBL}\includegraphics[width=0.33\linewidth]{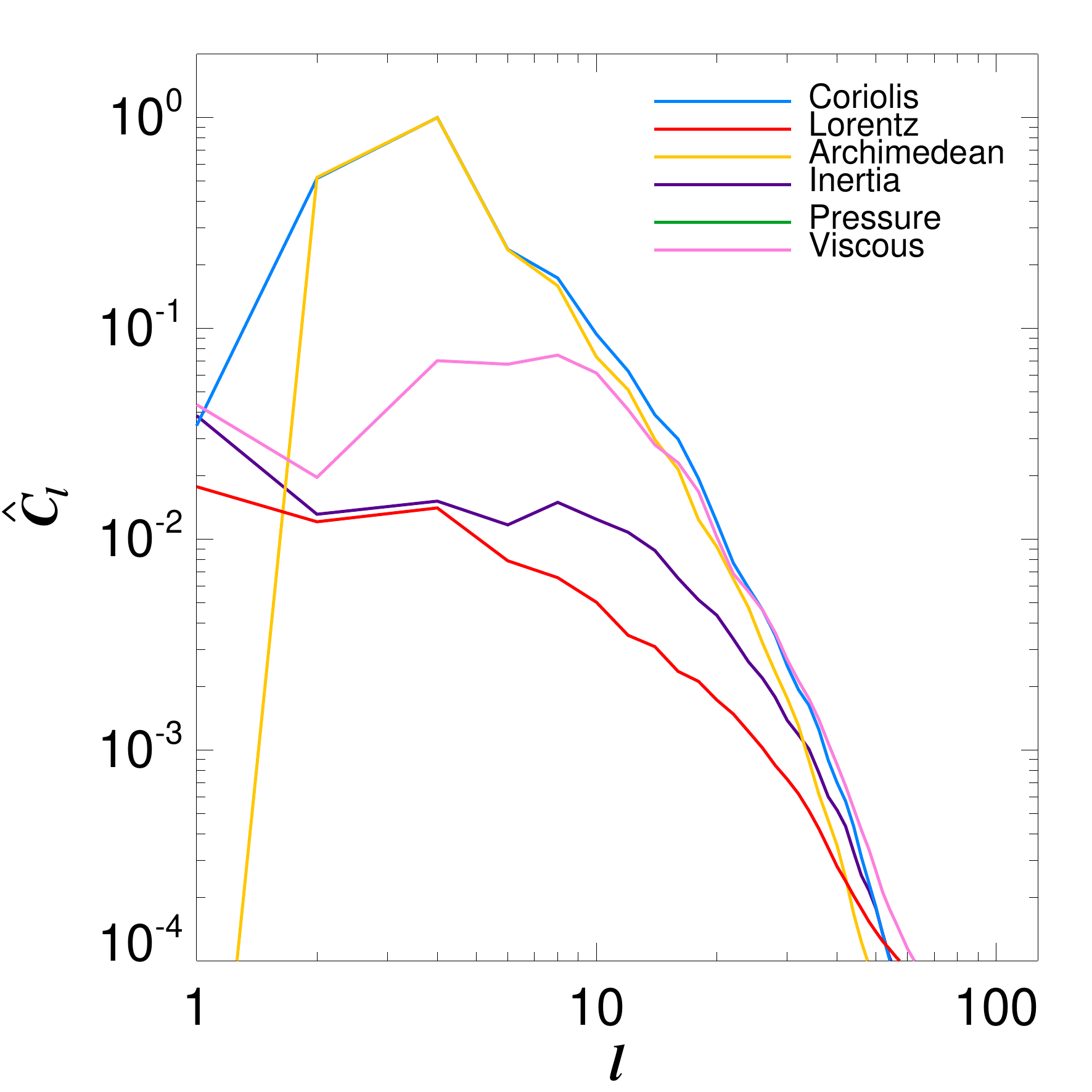}}
\hspace{1mm}
(c)\hskip -5mm{\label{fig:Pm12_2.05Rac_PlnoBL}\includegraphics[width=0.33\linewidth]{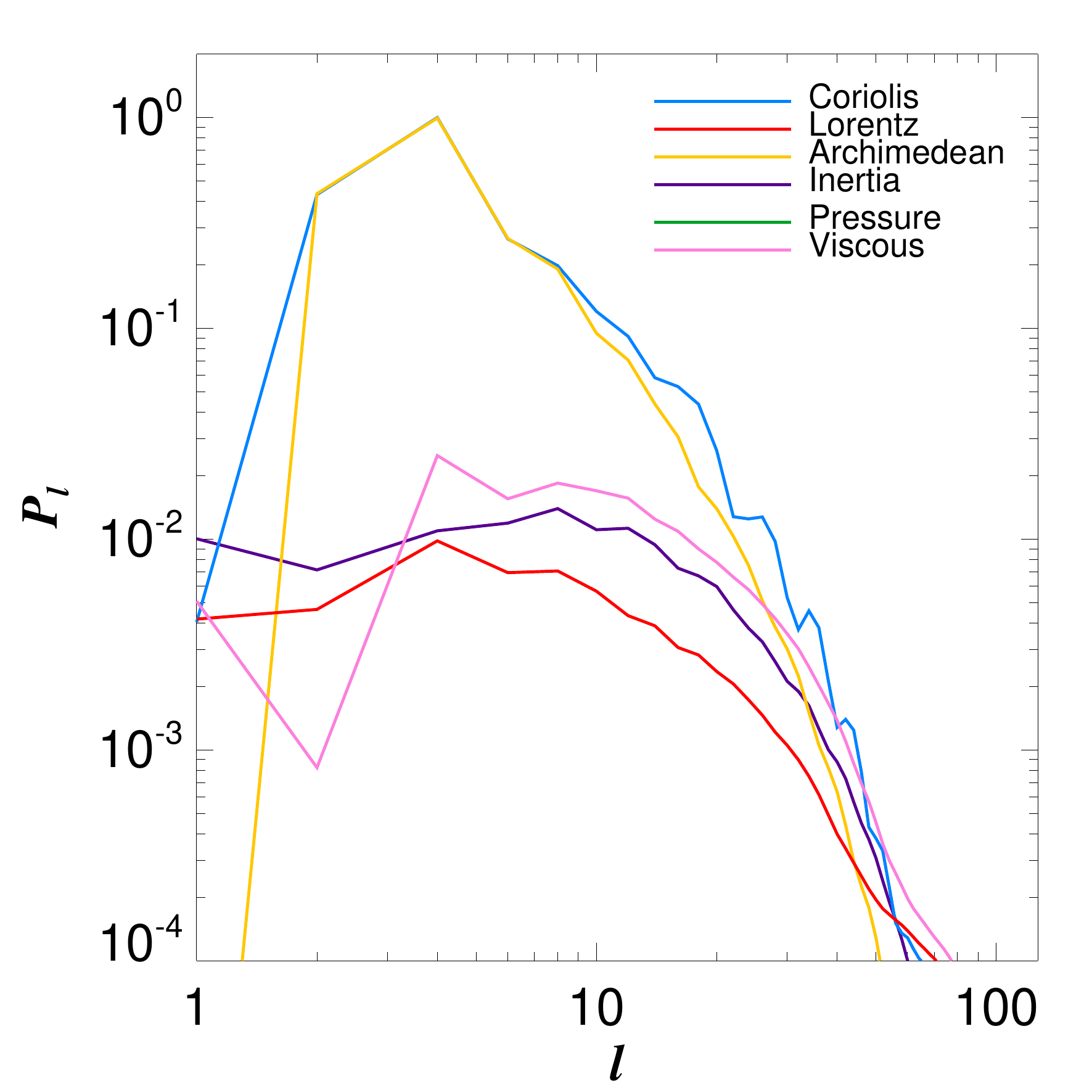}}}
\centerline{
(d)\hskip -5mm{\label{fig:Pm12_10Rac_FlnoBL}\includegraphics[width=0.33\linewidth]{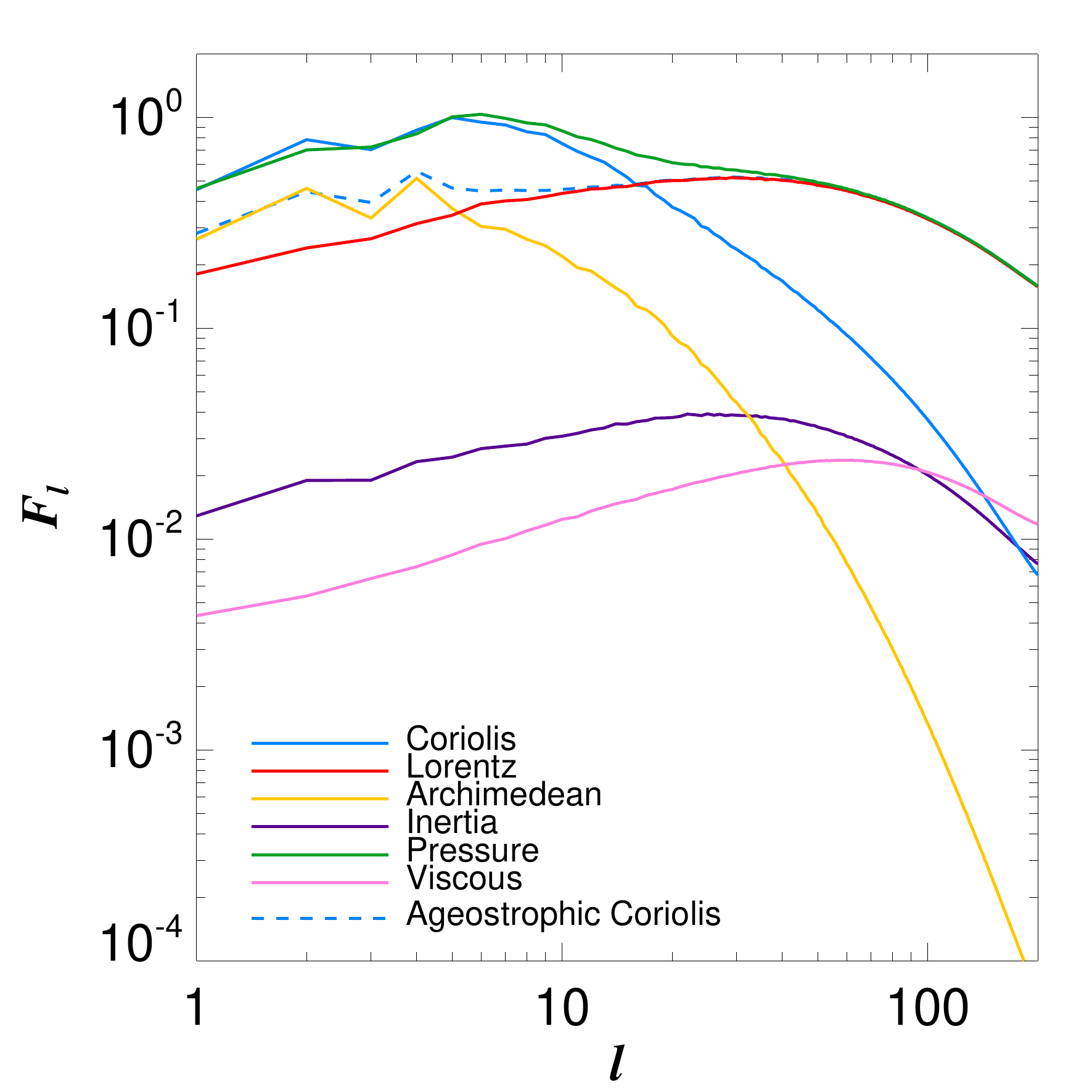}}
\hspace{1mm}
(e)\hskip -5mm{\label{fig:Pm12_10Rac_ClnoBL}\includegraphics[width=0.33\linewidth]{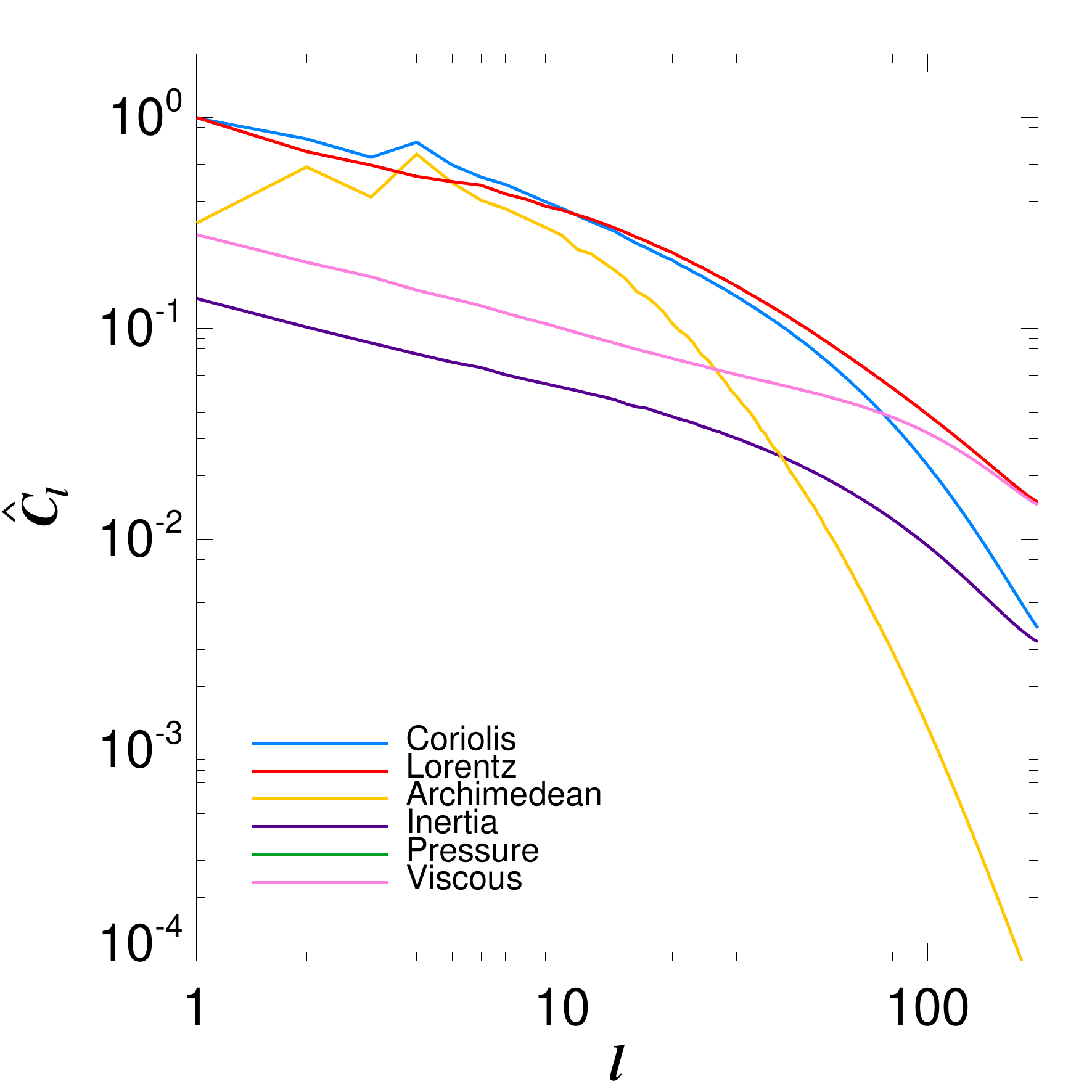}}
\hspace{1mm}
(f)\hskip -5mm{\label{fig:Pm12_10Rac_PlnoBL}\includegraphics[width=0.33\linewidth]{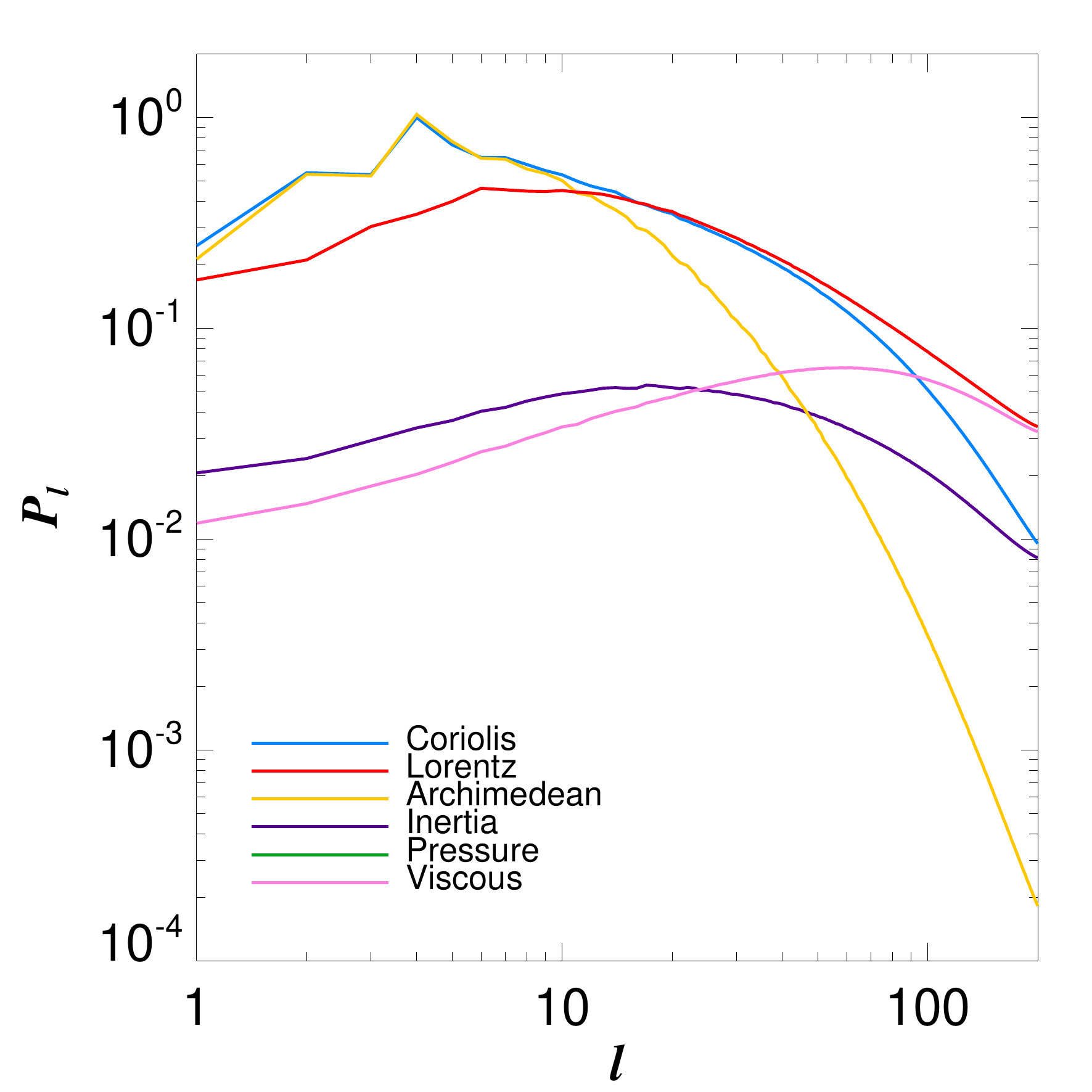}}}
\centerline{
(g)\hskip -5mm{\label{fig:Pm1_30Rac_FlnoBL}\includegraphics[width=0.33\linewidth]{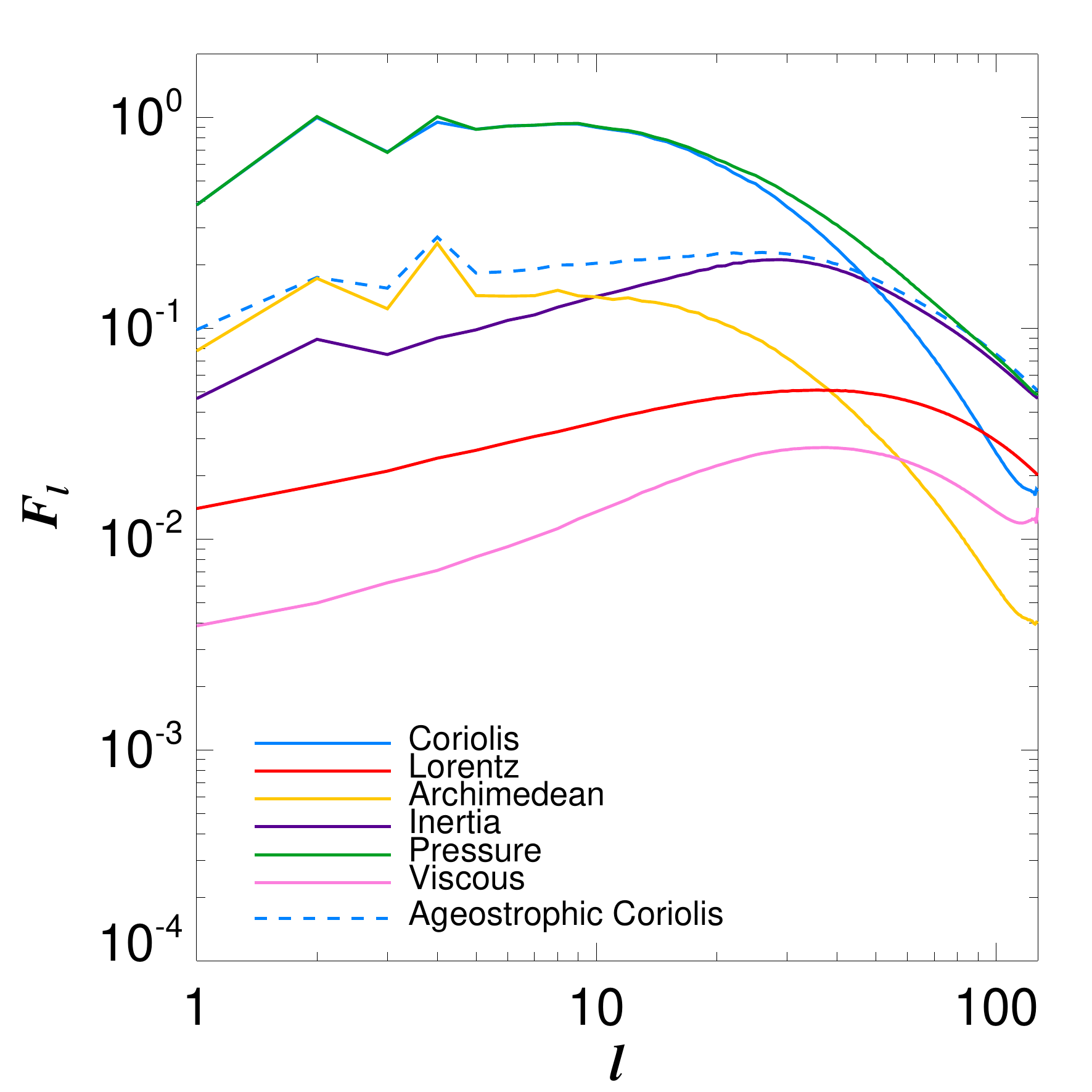}}
\hspace{1mm}
(h)\hskip -5mm{\label{fig:Pm1_30Rac_ClnoBL}\includegraphics[width=0.33\linewidth]{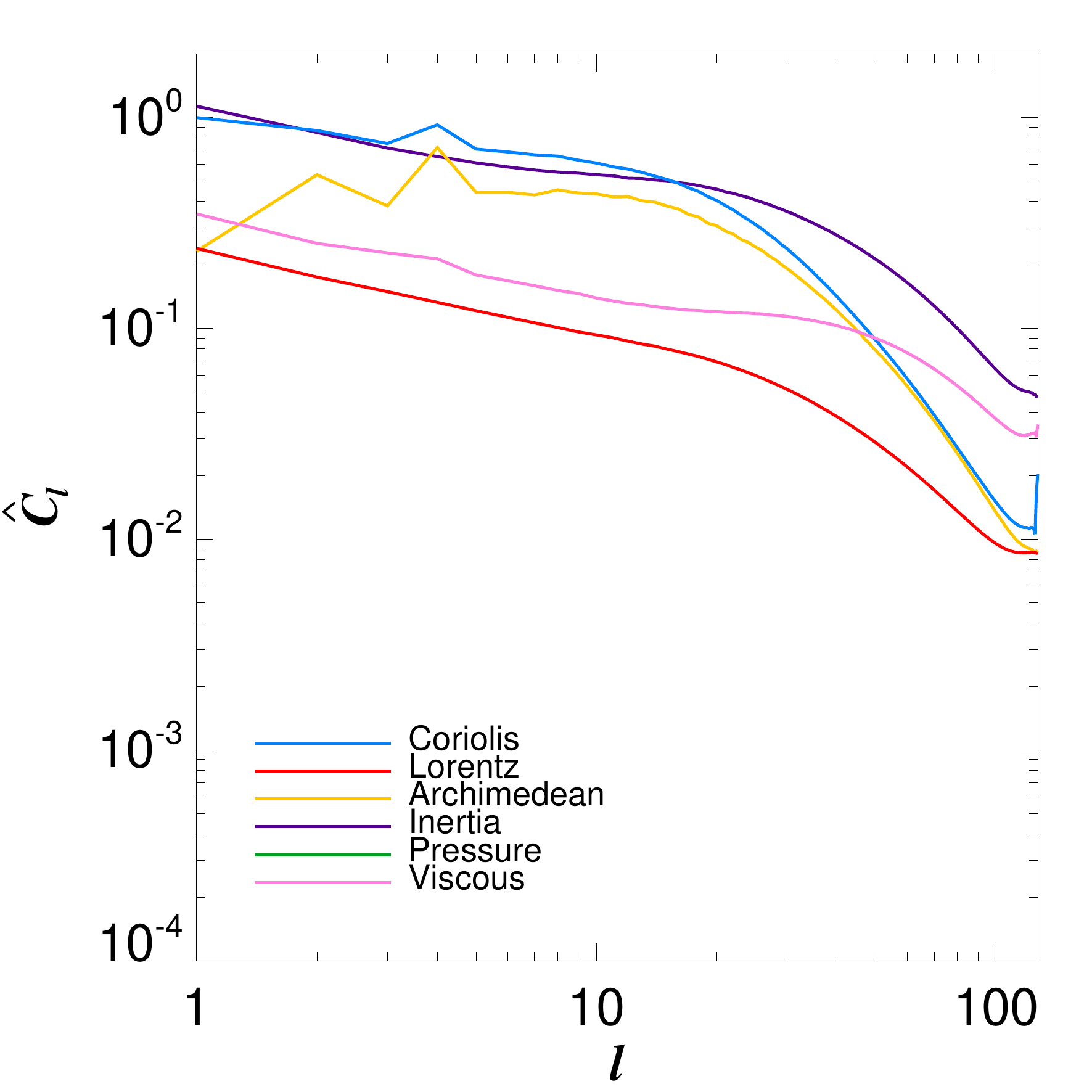}}
\hspace{1mm}
(i)\hskip -5mm{\label{fig:Pm1_30Rac_PlnoBL}\includegraphics[width=0.33\linewidth]{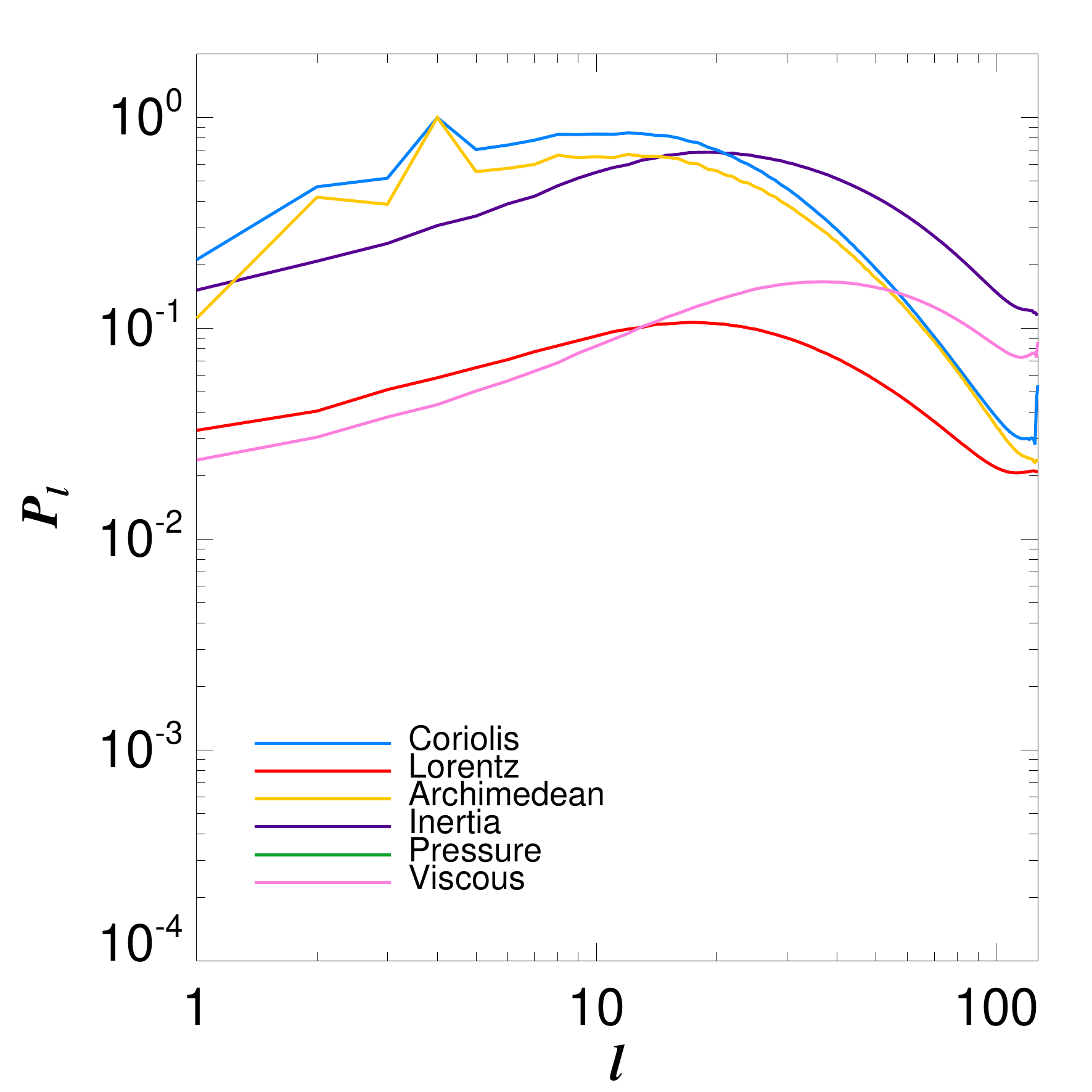}}}
\caption{Force spectra representations for case A (a,b,c) %($E=10^{-4}$, $Ra=2.05Ra_c$, $Pm=12$), 
for which odd modes (except $l=1$) are not plotted due to strong symmetry about the equator,
case B (d,e,f)
and case C (g,h,i).
% ($E=10^{-4}$, $Ra=10Ra_c$, $Pm=12$).
}
\label{fig:rundyn}
\end{figure}

Fig.~\ref{fig:rundyn}a,b,c, for case A, predictably shows that the Lorentz force enters only as an additional secondary force at all scales and never enters the leading order balance of solenoidal forces. The role of the viscous force is not obvious in $F_l$ but is clearly revealed in the solenoidal forces; in both $\hat{C}_l$ and $P_l$, $\forcevec{V}$ enters the primary balance over a wide range of scales demonstrating that this dynamo model is strongly influenced by viscous effects, contrary to what Fig~\ref{fig:rundyn}a indicates.
The balance is AC and then VAC beyond $l\sim10$; as expected for such balances, the resultant flow is strongly $z$-independent (Fig.~\ref{fig:mersecs10}b). 

%% SD discussion
Fig.~\ref{fig:rundyn}d,e,f, %for $Pm=12$,
for case B
%the SD run,
is typified by the Lorentz force entering the primary balance of forces.
Fig.~\ref{fig:rundyn}d, for $F_l$, is similar to plots found in previous work \citep{schwaiger2019}.
Whilst the importance of the Lorentz force can be observed in $F_l$, the picture is again
different when the solenoidal forces are formed with plots of $\hat{C}_l$ and $P_l$ (Figs.~\ref{fig:rundyn}e,f) 
revealing a MAC balance for $l\lesssim10$, beyond
which an MC balance prevails for %a range of intermediate lengthscales
$10\lesssim l\lesssim100$. 
The solenoidal forces again show the viscous force entering 
the primary balance at small scales.
In contrast to previous runs (where it entered either across a wide range of lengthscales or in a small scale VI balance), here the balance is VM for $l\gtrsim100$. 
The effect of the change to a magnetically-influenced balance can be seen in a meridional section of the flow
(Fig.~\ref{fig:mersecs10}c). The structure of the
flow is no longer vertically aligned, demonstrating departure from
rotationally-dominated flow. 
The Lorentz force relaxes the Taylor-Proudman constraint
in a way that it (and neither inertial nor viscous forces) was unable to
do in the previous regimes.
Because the Leray projector suffers from a non-uniqueness due to the harmonic field, we place the emphasis  on Fig.~\ref{fig:rundyn}e, which is uniquely defined, and highlights both the primary importance of the Lorentz force at all scales and the irrelevance, from a dynamical point of view, of curve crossings in Fig.~\ref{fig:rundyn}d.

\begin{figure}
\centering
(a)\hskip -5mm{\label{fig:Pm1_10Rac_FlnoBL}\includegraphics[width=0.33\linewidth]{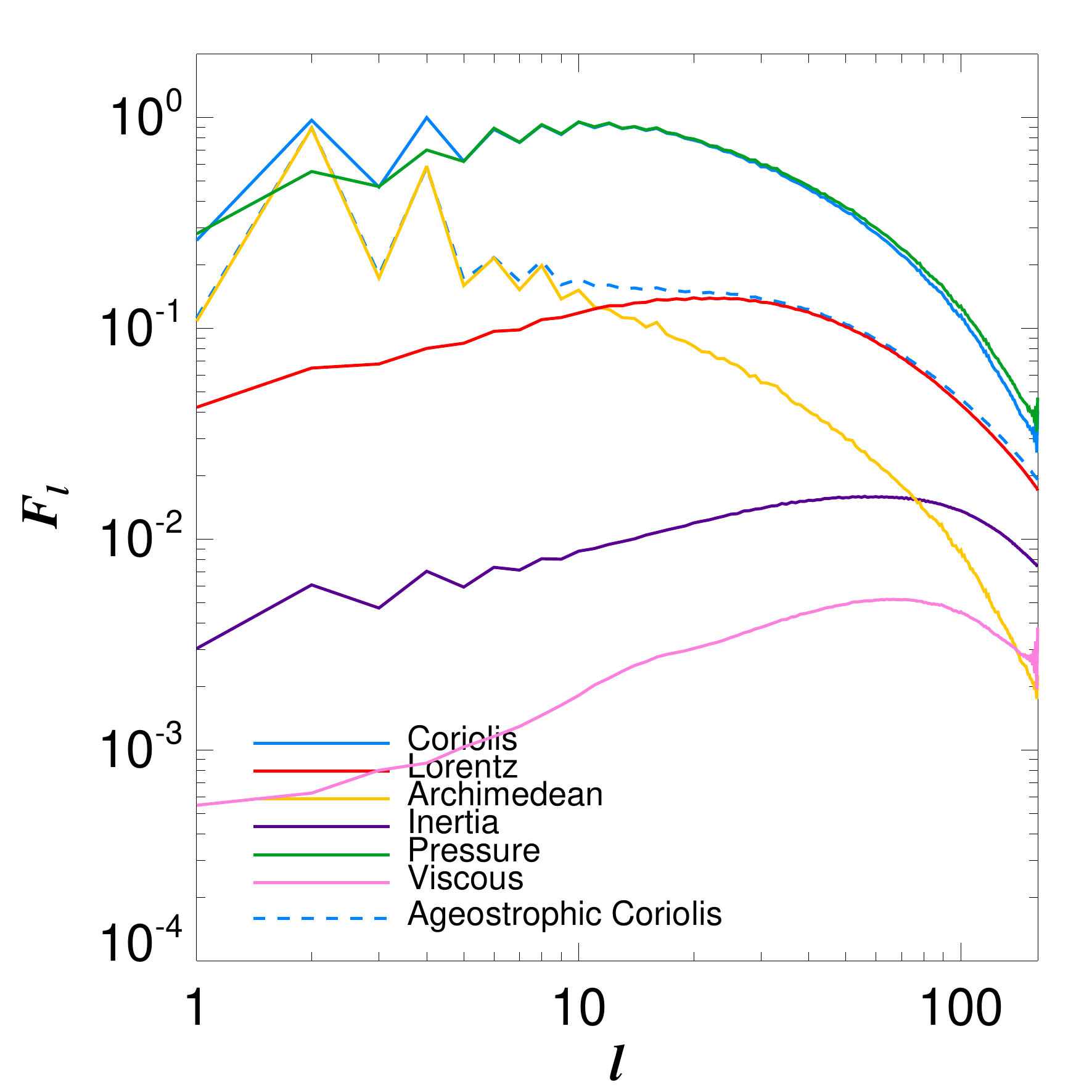}}
\hspace{1mm}
(b)\hskip -5mm{\label{fig:Pm1_10Rac_ClnoBL}\includegraphics[width=0.33\linewidth]{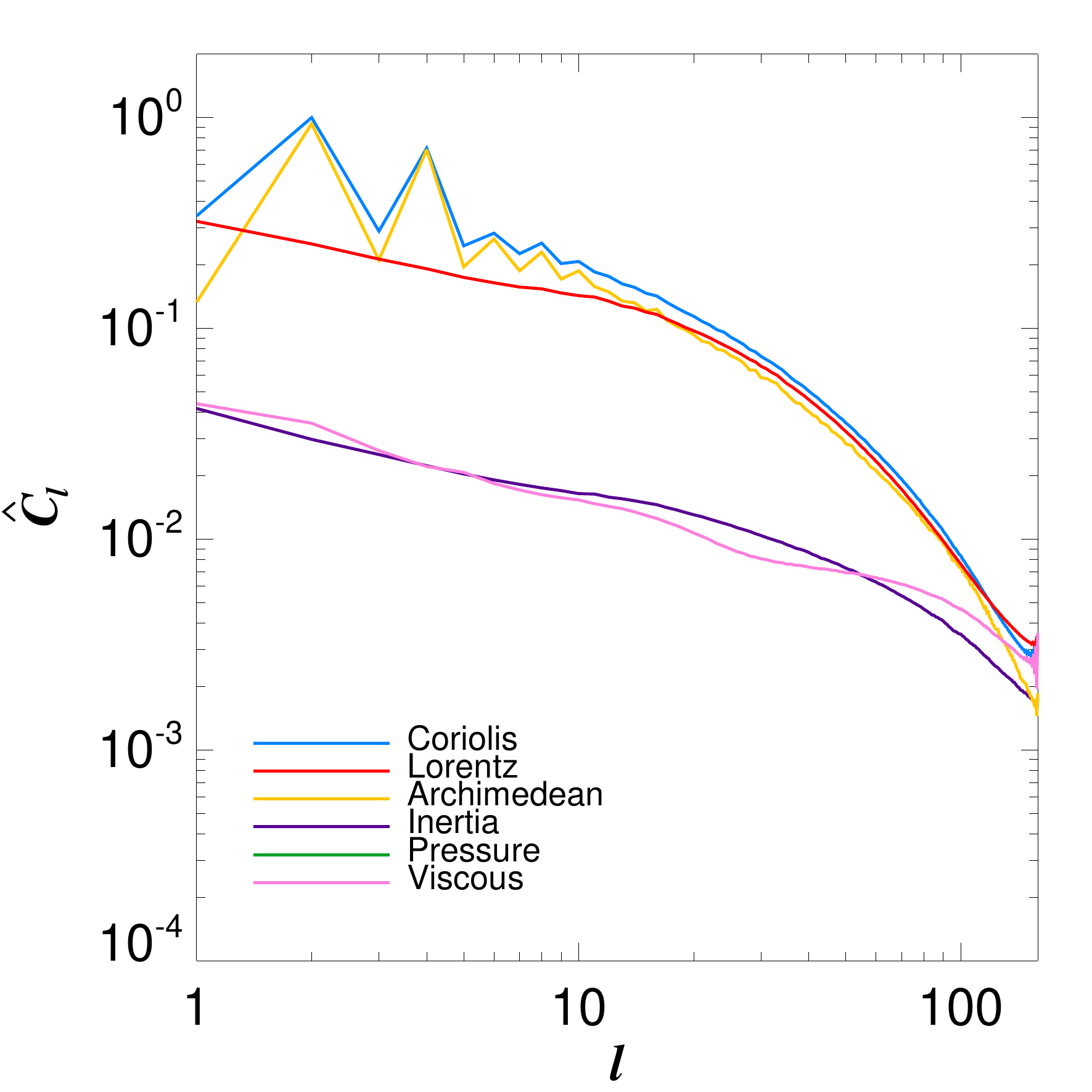}}
\hspace{1mm}
(c)\hskip -5mm{\label{fig:Pm1_10Rac_PlnoBL}\includegraphics[width=0.33\linewidth]{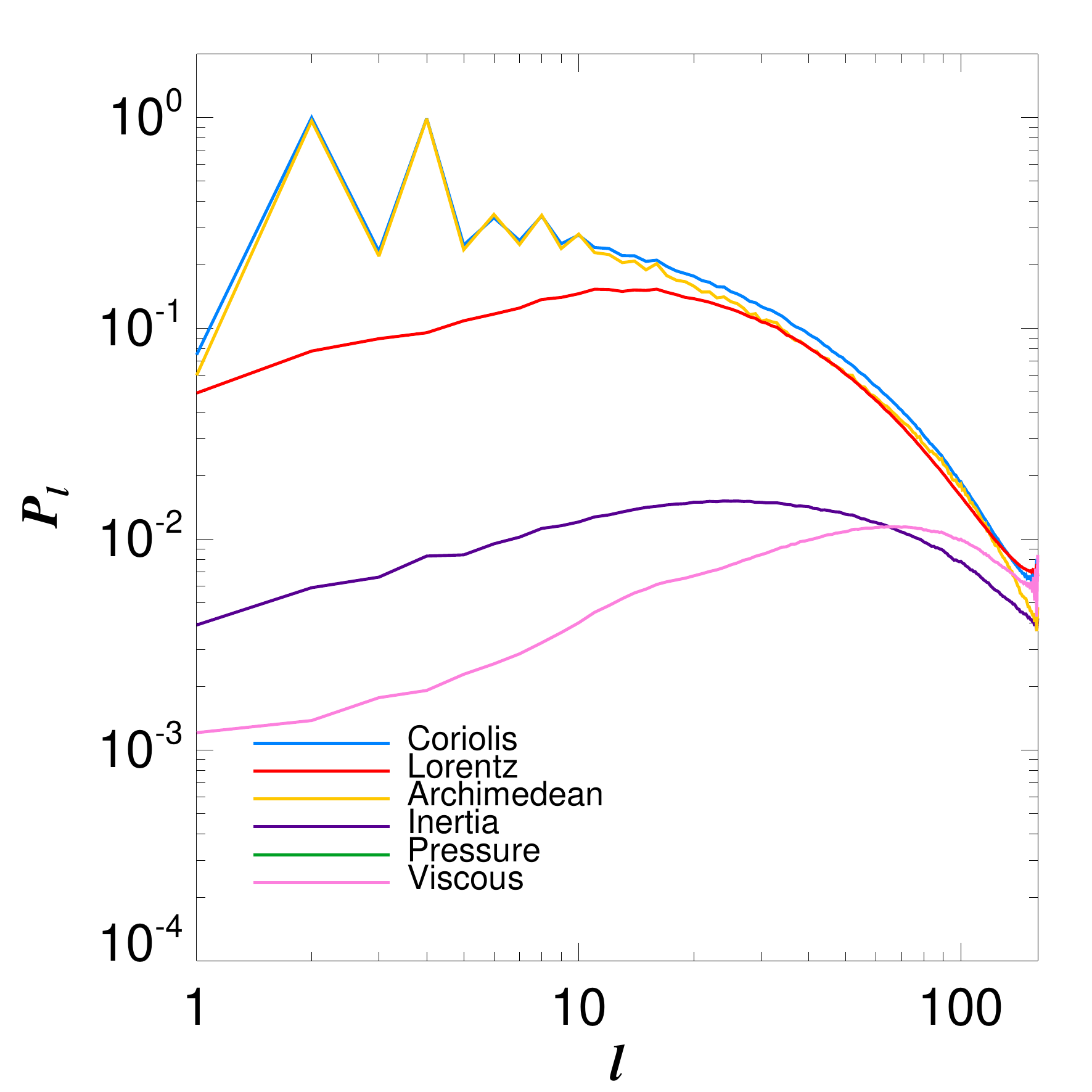}}
%\vspace{-10mm}
\caption{Force spectra representations for case D.
}
\label{fig:E1e-5}
\end{figure}

A multipolar dynamo regime is obtained with sufficiently strong driving. Inertia becomes increasingly important over a range of lengthscales; as shown in Fig.~\ref{fig:rundyn}g,h,i, for case C
%the FM run
, the contribution from the inertial
force at all lengthscales is much greater compared with the less strongly driven cases.
%weak dipolar field case.
Plots of $F_l$ show the inertial force entering
the zeroth order balance at small scales; %in balance with the pressure gradient and ageostrophic Coriolis force.
\cite{schwaiger2019} exhibit a
similar plot to Fig.~\ref{fig:rundyn}g for a multipolar simulation
at alternative input parameters. However, in Figs.~\ref{fig:rundyn}h,i solenoidal force balances reveal the inertial force entering at leading order across all lengthscales and it becomes the dominant solenoidal force for $l\gtrsim20$. This again demonstrates the 
potential pitfall in
%inaccuracy of
considering the ageostrophic Coriolis force over a more complete approach that removes gradients fully. %, such as curled or projected forces. 
The Lorentz force is weak and does not enter the leading order balance at
any scale. %for this multipolar run.
We finally report a simulation (case D)
at $E=E_\eta=10^{-5},$ and a forcing of 10 times critical; Fig.~\ref{fig:E1e-5}
shows force balances for this simulation. Fig.~\ref{fig:E1e-5}a is, in many respects, similar to some of the force balances published in the literature for state-of-the-art simulations. A clear crossing is present between the buoyancy and the Lorentz force curves. However, it does not correspond to a dynamically relevant lengthscale, as revealed by the curl of forces in Fig.~\ref{fig:E1e-5}b. In this simulation the solenoidal parts of the buoyancy, Lorentz, and Coriolis forces appear to be in general balance, though with some fluctuations. Further work is clearly needed to ascertain whether a more accurate
balance of Coriolis and Lorentz forces at all scales (as seen in Fig.~\ref{fig:rundyn}e) can be achieved by slightly decreasing $E_\eta$ at this value of $E$ with fixed forcing.

%%%%%%%%%%%%%%
\section{Discussion}

We have presented three different representations of
force balances in several spherical dynamo simulations.
Our length-dependent plots of traditional forces ($F_l$)
are similar to those found previously
\citep{aubert2017,schwaiger2019,schwaiger2021,gillet2021}.
Broadly speaking, $F_l$ is
dominated by a zeroth order geostrophic balance 
at least at large scales.
It can also be argued that,
at large scales, a first order AC$^\textrm{ag}$ balance then exists.
In case B/C
%the SD/FM regime
the Lorentz/inertial force also becomes
significant at zeroth order. 
For case B, 
%the SD regime,
the claim could be that a zeroth order geostrophic balance at large scales gives way to a zeroth order magnetostrophic balance at smaller scales; however,
it is unclear what useful meaning
%this wrongly assumes
$\FCag$ retains once geostrophy is broken.
The remaining
forces: inertia
(in the HD regime and cases A, B, and D),
%(in HD, WD, and SD regimes),
Lorentz 
(in cases A and C),
%(in WD and FM regimes), 
and viscous (in all regimes), appear to be secondary forces across all scales.
In particular, the viscous
force is always a secondary force in each regime under this representation
of the forces. 
%However, this is not numerically realistic since the
%viscous force is needed from a numerical point of view to regularise the smallest scales. 
{
However, this is numerically puzzling
as the resolution using a spectral method 
requires small-scale regularity. The
viscous force is thus needed from a numerical point of view to regularise the smallest scales. 
}

Removing the gradient part of each force (including, consequently, the whole pressure gradient force and zeroth order balance)
to form `solenoidal forces' ($\hat{C}_l$ and $P_l$) is desirable because it allows direct
access to the balance of forces that controls the dynamics. 
$P_l$ has a degree of arbitrariness in the choice of the harmonic field, whereas $\hat{C}_l$ is uniquely defined.
The first order balance in $\hat{C}_l$ and $P_l$ is AC at large scales with the Lorentz/inertial force also entering in 
case B/C.
%the SD/FM regime. 
At smaller scales (near the tail of the spectra), the viscous force
now enters the balance in each simulation.
When the zeroth order balance is geostrophic, the leading order balance in $\hat{C}_l$ and $P_l$ essentially recovers the first order balance shown in $F_l$.
This is because, in such situations, the solenoidal part of the Coriolis force is well approximated by its
ageostrophic part; in other words
$\mathbb{P}(\forcevec{C})\approx\FCag \, .$ 
However,
as $l$ increases and $\forcevec{C}$ loses its predominance in the
zeroth order balance, its ageostrophic part becomes an unreliable
approximation for its solenoidal part. 
We found that 
$\mathbb{P}(\forcevec{C})\approx\FCag$ was only reliably satisfied for $l\lesssim 10$; it is a particularly
poor measure in 
case B
%the SD regime
where the Lorentz force enters the
first order balance even at the largest scales. 
Notably, this regime
is the most relevant for the geodynamo; for this reason
%the ageostrophic Coriolis force should be avoided in favour
we advocate the use of solenoidal forces, which provide an accurate measure for the dynamically relevant balance across all scales. 
\cite{schwaiger2019,gillet2021} argue that important lengthscales for the flow can be
derived from the crossings of $F_l$ curves.
{We argue that such crossings of terms containing gradient parts can be misleading when trying to understand the dynamics of an incompressible (Boussinesq) flow.} If such crossings are relevant, they should be observed on the curl of forces instead.
It is also worth nothing that, in addition to the scale dependence we have presented, force balances may also depend on position; for example, between the regions inside and outside the tangent cylinder. A spectral decomposition, by its nature, cannot capture such inhomogeneity in position.

We constructed two representations of solenoidal forces. Some authors also compare the azimuthal average of the azimuthal component of force balances in order to be free of any gradient effects (this is the case for example in \cite{Sheyko2018, Menu2020})
The advantage of curls is there is no arbitrariness in the
calculation; gradient parts are eliminated in a unique manner. The potential disadvantage of taking curls is their tendency to
inflate sharp spatial changes.
It is possible to partly compensate $C_l$ to form $\hat{C}_l$;
however, this does not account for radial derivatives
introduced when curling.
Projecting the forces onto their solenoidal part has the advantage of no extra derivative. However, a harmonic field is introduced
adding a degree of arbitrariness
to $P_l$. In our work, the chosen condition on this field was
that each force has vanishing radial part on the boundaries. 
Nevertheless, although plots of $\hat{C}_l$ and
$P_l$ may appear quantitatively different at large scales, they present the same
qualitative picture and are equivalent at small scales. In particular, either representation
offers a clearer picture of the balance controlling the flow than the
comparison of full forces.

\backsection[Acknowledgements]{The authors are grateful to Steve Tobias for discussions. This work was supported by DiRAC Project ACTP245. This work was presented at the Isaac Newton Institute in Cambridge, UK on 16 May 2022 (ED) and on 14 Sept.~2022 (RT).}

\backsection[Declaration of interests]{The authors report no conflict of interest.}

%% BIBLIO
\bibliographystyle{jfm}
\bibliography{References}

\end{document}